\documentclass[prl,twocolumn,showpacs,superscriptaddress,10pt]{revtex4-1}
\usepackage{hyperref}
\usepackage{graphicx}
\usepackage{epsfig}
\usepackage{subfigure}
\usepackage{setspace}
\usepackage{amsmath}
\newcommand{\mum}{~\mu \mathrm{m}}

\begin{document}

\title{Polarization degenerate micropillars fabricated by designing elliptical oxide apertures}

%\author{author1}
\author{Morten P. Bakker}
\affiliation{Huygens-Kamerlingh Onnes Laboratory, Leiden University, P.O. Box 9504, 2300 RA Leiden, The Netherlands}
\author{Ajit V. Barve}
\affiliation{University of California Santa Barbara, Santa Barbara, California 93106, USA}
\author{Alan Zhan}
\affiliation{University of California Santa Barbara, Santa Barbara, California 93106, USA}
\author{Larry A. Coldren}
\affiliation{University of California Santa Barbara, Santa Barbara, California 93106, USA}
\author{Martin P. van Exter}
\affiliation{Huygens-Kamerlingh Onnes Laboratory, Leiden University, P.O. Box 9504, 2300 RA Leiden, The Netherlands}
\author{Dirk Bouwmeester}
\affiliation{Huygens-Kamerlingh Onnes Laboratory, Leiden University, P.O. Box 9504, 2300 RA Leiden, The Netherlands}
\affiliation{University of California Santa Barbara, Santa Barbara, California 93106, USA}

\date{\today}

\begin{abstract}

A method for fabrication of polarization degenerate oxide apertured micropillar cavities is demonstrated.
Micropillars are etched such that the size and shape of the oxide front is controlled.
The polarization splitting in the circular micropillar cavities due to the native and strain induced birefringence can be compensated by elongating the oxide front in the [110] direction, thereby reducing stress in this direction.
By using this technique we fabricate a polarization degenerate cavity with a quality factor of 1.7 $\times 10^4$ and a mode volume of $~2.7 \mum^3$, enabling a calculated maximum Purcell factor of 11.

\end{abstract}

%\pacs{73.20.Mf, 42.25.Dd, 78.66.-n, 42.30.Ms}

\maketitle

Quantum dots in micropillar cavities form an interesting platform for cavity quantum electrodynamics experiments in the solid state \cite{Reitzenstein2010}.
For example, the coupling between the spin in a singly-charged quantum dot (QD) and the polarization of a photon in the Purcell regime holds promise for applications in hybrid quantum information processing \cite{Bonato2010}.
For this purpose oxide apertured micropillars are attractive as they combine simple fabrication of voltage contacts, excellent mode-matching with external fields and access to the Purcell regime\cite{Stoltz2005}\cite{Rakher2009}.

An important challenge however is to obtain polarization degenerate cavity modes.
This is an important condition in order to prepare entanglement between an electron spin and the polarization of a photon \cite{DeGreve2012}, which is essential for schemes as described in \cite{Bonato2010}.
Several post-processing techniques have been demonstrated to tune the polarization properties.
These techniques rely on the application of strain mechanically \cite{panajotov2000} or on the optical application of surface defects \cite{Bonato2009}\cite{VanDoorn1996}.
It is more desirable to obtain close-to polarization degenerate cavities after the wet oxidation processing step and thereby minimize the need of further tuning.

In this paper we demonstrate that, by systematically varying the shape of the etched micropillar, the shape of the oxide aperture can be controlled, thereby controlling the spectral properties.
Two arrays of micropillars of which the diameter and ellipticity are systematically varied are fabricated and the optical modes are characterized at 9.0 K.
Micropillars with circular oxide fronts exhibit optical modes with a large circular symmetry, but due to birefringence the fundamental mode is polarization non-degenerate.
However, for elliptical oxide fronts that are elongated in the [110] direction, the native birefringence is compensated for by strain-induced birefringence and polarization degenerate fundamental modes are obtained.

\begin{figure}[hb]
\centering
    \centerline{\includegraphics[angle=0]{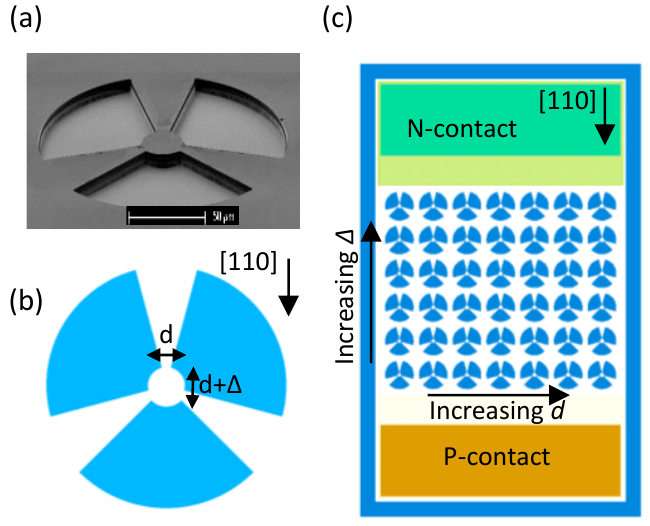}}
    \caption{(a) SEM image of a micropillar. (b) The micropillars have diameter $d$ and are elongated in the [110] direction by an amount $+\Delta$. (c) Systematic variations of $d$ and $\Delta$ are applied over an array.}
    \label{Fig1}
\end{figure}

The samples used in this study are grown by molecular beam epitaxy on a GaAs [100] substrate.
First a planar distributed Bragg reflector (DBR) cavity is grown which consists of a spacing layer and 26 pairs of GaAs/Al$_{0.90}$Ga$_{0.10}$As layers in the top mirror and 29 pairs in the bottom mirror.
%The imbalance in the DBR pairs is to compensate for the presence of a GaAs substrate.
The spacing layer consists of a $\lambda$ GaAs layer, containing a layer of InAs self-assembled QDs in the center \cite{petroff2001}, and a $\frac{3}{4} \lambda$ Al$_x$Ga$_{1-x}$As aperture region.
The oxidation aperture consists of a 10 nm AlAs layer embedded between 95 nm Al$_{0.83}$Ga$_{0.17}$As and 66 nm Al$_{0.75}$Ga$_{0.25}$As layers, providing a linearly tapered oxidation upon the wet oxidation.
Micropillars are etched such that they are connected to the bulk region via three bridges, to provide global electrical contacts, as shown in Fig. \ref{Fig1} (a).

This geometry was found to be an optimum as for two bridges the oxide front is found to be more elliptically shaped, while for more than three bridges the bridges are too thin and the risk increases that the electrical conductance to the micropillar center is insufficient \cite{Strauf2007}.
This geometry, together with the etching process, is however expected to induce in-plane anisotropic strain, which needs to be compensated for together with the native birefringence, that can be present even in perfectly circular mesas.

Figures \ref{Fig1} (b,c) schematically show that the micropillar diameter $d$ in the $[1\bar{1}0]$ direction and diameter $d$ $+\Delta$ in the [110] direction are systematically varied in a 6 $\times$ 7 array, with $d$ = [29, 30, 30.5, 31, 31.5, 32, 33] $\mu$m and $\Delta$ = [0, 0.5, 1.0, 1.5, 2.0, 2.5] $\mu$m.
Then a wet thermal oxidation procedure to form an oxide aperture is applied \cite{Baca2005}.
Finally electrical contacts to the p-doped and n-doped GaAs surrounding the QDs are fabricated.

To characterize the optical properties of the confined optical modes, standard microphotoluminescence techniques are used.
The sample is held in a cryostat at 9.0 K and pumped using an 852 nm laser diode to excite QD emission.
We characterize the anisotropy of every micropillar in two different ways.
First of all, we measure the polarization-splitting of the fundamental mode to characterize the birefringence.
Second, we measure the wavelength differences between the first-order transverse modes and the fundamental mode.
This transverse mode splitting is linked to the optical confinement, which can be different in the two directions.

Finally, in order to get an indication of the shape of the buried oxide aperture layer, which determines the optical confinement, a spatial reflection scan is performed at room temperature.
For this, a laser with a wavelength $\lambda = 1064$ nm located outside of the DBR stopband is used, such that the reflectance depends on interfering reflections from the top and bottom DBR mirrors.
This interference is a function of the optical length of the spacing layer and therefore the reflectance depends on the buried oxide thickness \cite{Bakker2013}.
The front of the oxide is clearly visible as a ring with a lower reflectivity.

\begin{figure}[h]
\centering
    \centerline{\includegraphics[angle=0]{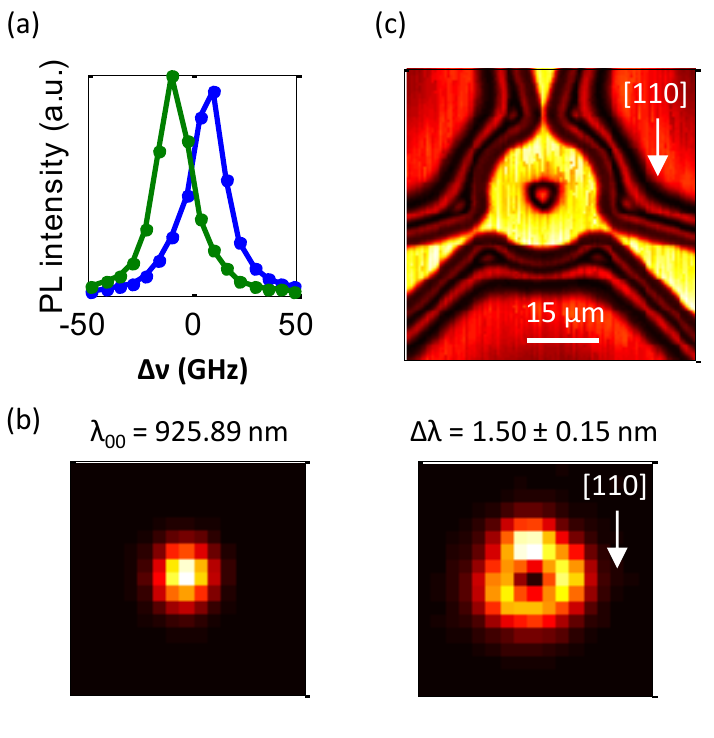}}
    \caption{(a) PL from the fundamental $\Psi_{00}$ mode at two orthogonal polarizations. (b) Spatial PL scans at the wavelength of the fundamental mode and a wavelength interval overlapping with both first-order modes. Wavelengths are selected using a spectrometer. (c) Spatial reflectivity scan of a focused $\lambda$ = 1064 nm laser spot indicates the circular oxide front.}
    \label{Fig2a}
\end{figure}

\begin{figure}[h]
\centering
    \centerline{\includegraphics[angle=0]{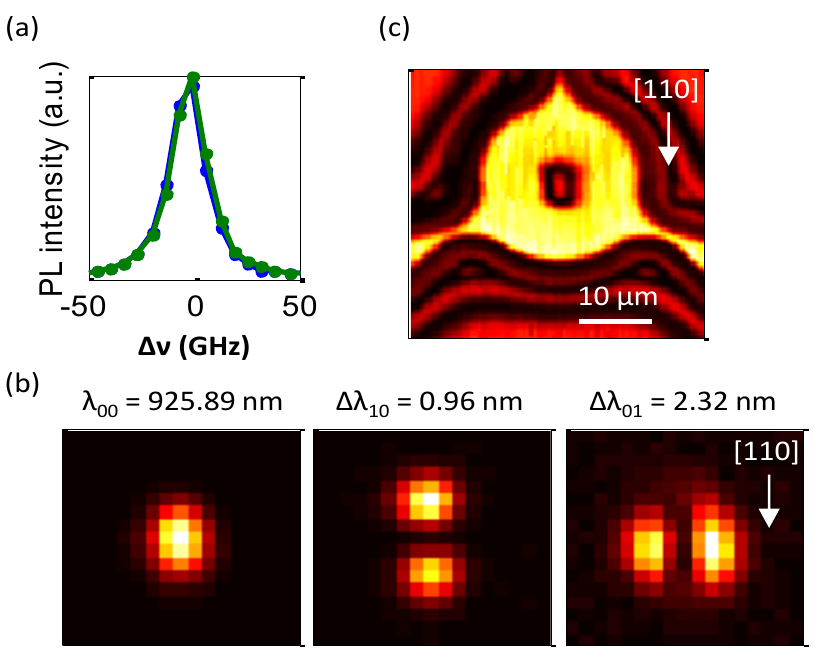}}
    \caption{(a) PL from a polarization degenerate $\Psi_{00}$ cavity mode. (b) Spatial PL scans of the fundamental $\Psi_{00}$ and the first-order $\Psi_{10}$ and $\Psi_{01}$ modes. (c) Spatial $\lambda$ = 1064 nm reflectivity scan that indicates the oxide front is elongated in the [110] direction.}
    \label{Fig2b}
\end{figure}

Figure \ref{Fig2a} shows a micropillar that was elongated slightly, by 0.5 $\mu$m, in the [110] direction.
Due to a faster wet oxidation rate in this direction this results in a nearly circular oxidation front as shown in Fig. \ref{Fig2a} (c).
The circular symmetry is apparent as well in the spatial profiles of the confined modes in Fig. \ref{Fig2a} (b), where the incoherent sum of the two first higher-order Hermite-Gaussian modes resemble a Laguerre-Gaussian transverse mode profile.
In Fig. \ref{Fig2a} (a) however a clear frequency splitting between two linear orthogonal polarization modes of the fundamental mode is visible due to birefringence.

Figure \ref{Fig2b} shows an even more elliptical micropillar, elongated in the [110] direction by 2 $\mu$m.
This elongated shape is now also visible in the shape of the buried oxide aperture in Fig. \ref{Fig2b} (c).
In Fig. \ref{Fig2b} (b) clear Hermite-Gaussian modes are identified that now posses a great difference between the modesplittings $\Delta\lambda_{10(01)} = \lambda_{00}-\lambda_{10(01)}$, owing to a difference in the amount of optical confinement in orthogonal directions.
We define $\Delta\lambda_{10}$ to be in the [110] direction.
The polarization splitting of the fundamental mode however is about 1 GHz, less than 6\% of the FWHM, indicating the birefringence has been strongly reduced.

\begin{figure}[h]
\centering
    \centerline{\includegraphics[angle=0]{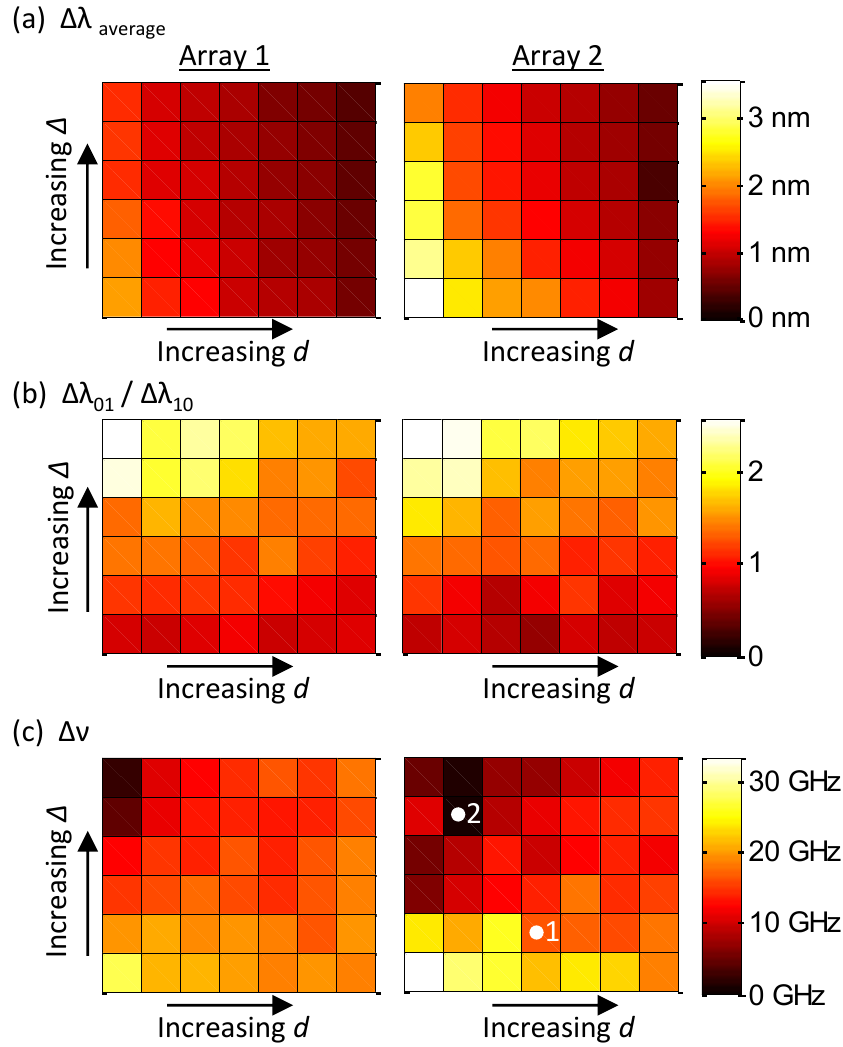}}
    \caption{Colormaps indicate from two arrays: (a) the average modesplittings $\Delta\lambda_{average}$ between the first order $\Psi_{10} / \Psi_{01}$ modes and the fundamental $\Psi_{00}$ mode, (b) the ratio $\Delta\lambda_{01}/\Delta\lambda_{10}$ of the modesplittings between the $\Psi_{01} / \Psi_{10}$ and the $\Psi_{00}$ modes, and (c) the polarization splitting $\Delta\nu$ of the $\Psi_{00}$ mode. Array 2 exhibits a larger average modesplitting and thus has oxidized slightly further. The dots 1 and 2 denote the cavities displayed in Fig. \ref{Fig2a} and Fig. \ref{Fig2b}, respectively.}
    \label{Fig3}
\end{figure}

Figure \ref{Fig3} shows the result of a systematic characterization of two arrays, of which array 2 is oxidized slightly further.
Figure \ref{Fig3} (a) shows the modesplittings between the fundamental and first order modes, averaged over the two linear polarizations.
Clearly, the average modesplitting decreases as the size of the micropillar is increased, as expected.
Figure \ref{Fig3} (b) shows the ratio $\Delta\lambda_{01}/\Delta\lambda_{10}$ between the modesplittings in two directions.
An increasing ratio $\Delta\lambda_{01}/\Delta\lambda_{10}$ corresponds to less optical confinement in the [110] direction with respect to the [1$\bar{1}$0] direction which arises from a more elongated oxide front.
This correlates with increasing $\Delta$.
Figure \ref{Fig3} (c) displays the polarization splitting $\Delta\nu$ of the fundamental $\Psi_{00}$ mode.
A clear relation is visible between $\Delta\lambda_{01}/\Delta\lambda_{10}$ and $\Delta\nu$.

We qualitatively explain our findings by a modification of the birefringence under the influence of uniaxial strain in the [110] direction \cite{VanDoorn1996}.
Even for the almost circular apertures that remains after oxidation, some uniaxial strain is expected as the oxide layer is more extended in the [110] direction.
This can be the result of an anisotropy of the oxidation rate in combination with the location of the three bridges.
When the oxide front is more elongated however we expect the strain to be reduced such that the birefringence can be fully compensated for.

An important figure of merit of microcavities is the Purcell factor.
For the cavity shown in Fig. \ref{Fig2b} we find a $Q$-factor of $Q \approx 1.7 \times 10^4$ and by following methods described in \cite{Bonato2012} we predict that a maximum Purcell factor of 11 can be achieved.

%We would like to remark that shape birefringence due to anisotropic confinement is small ($<$ 1 GHz) in our %micropillars as vector diffraction effects are small.
%Therefore a paraxial (and scalar) description suffices \cite{SnyderLove}.

We would like to remark that shape birefringence due to anisotropic confinement is small ($<$ 1 GHz) in our micropillars and a paraxial (and scalar) description practically suffices.
The polarization/vector correction to the scalar wave equation, as calculated from perturbation theory \cite{SnyderLove}\cite{Weisshaar1995}, predicts that the shape-induced birefringence of the fundamental mode is a factor ($\Delta \lambda_{01} + \Delta \lambda_{10})/(2 \lambda_{00}$) smaller than the shape-induced confinement splitting ($\Delta \lambda_{01} - \Delta \lambda_{10}$), which results in a value of 0.84 GHz for the numbers mentioned in Fig. \ref{Fig2b}.

In conclusion we have shown it is possible to control the shape of the oxide aperture by the shape of the micropillar.
By applying systematic variations in the etched shapes, the strain-induced birefringence is varied and polarization degenerate cavities are obtained.
This is an appealing approach towards fabrication of polarization degenerate microcavities with minimal post-processing tuning techniques required.

We thank Thomas Ruytenberg for experimental assistance.
This work was supported by NSF under Grant No. 0960331 and 0901886 and FOM-NWO Grant No. 08QIP6-2.

\end{document}